\documentclass{aastex}
\usepackage{spr-astr-addons}
\usepackage{url}

\RequirePackage{color}

\begin{document}

\title{Mass-losing  accretion discs around supermassive black holes}
\shorttitle{Mass-losing discs}
\shortauthors{F. Khajenabi}

\author{F. Khajenabi\altaffilmark{1}}

\altaffiltext{1}{School of Physics, University College Dublin, Belfield, Dublin 4, Ireland\\ fazeleh.khajenabi@ucd.ie (FK)}

\begin{abstract}
We study the effects of outflow/wind on the gravitational stability of accretion discs around supermassive black holes using a set of analytical steady-state solutions. Mass-loss rate by the outflow from the disc is assumed to be a power-law of the radial distance and the amount of the energy and the angular momentum which are carried away by the wind are parameterized phenomenologically. We show that the mass of the first clumps at the self-gravitating radius {\it linearly} decreases with the total mass-loss rate of the outflow. Except for the case of small viscosity and high accretion rate, generally, the self-gravitating radius increases as the amount of mass-loss by the outflow increases. Our solutions show that as more angular momentum is lost by the outflow, then reduction to the mass of the first clumps is more significant.
\end{abstract}



%
\section{Introduction}

Accretion discs are believed to be present in a wide variety of
astronomical systems. The winds/outflows are potentially important as they may enhance  the accretion rate through carrying away angular momentum. It is now widely accepted that winds or outflows  have their origin in accreting systems, which at same time power the radiation emission associated with the object (e.g., Blandford \& Payne  1982; Fender, Belloni \& Gallo 2004). In a few cases, outflows have been demonstratively   observed in emission both from the broad line regime in narrow-line Seyfert 1 galaxies (e.g., Leighly \& Moore 2004; Leighly 2004) and from the narrow line region of Seyfert 1 galaxies (e.g., Das et al 2005; Das et al 2006). Outflows from Active Galactic Nuclei (AGN) are much more common than previously thought: the overal fraction of AGNs with outflows is fairly constant, approximately $60\%$, over many order of magnitude in luminosity (Ganguly \& Brotherton  2008).

The possible effects of winds/outflows on the radial structure of an accretion disc can be studied (e.g.,  Knigge 1999; Combet \& Ferreira 2008) within the framework of standard theory of accretion discs (Shakura \& Sunyaev 1973). It seems that outflows are more significant in the outer parts of a disc. But according to the standard theory of accretion discs (Shakura \& Sunyaev 1973), the outer parts of steady, geometrically thin, optically thick discs are prone to self-gravity and they might be expected to fragment into stars (e.g., Shlosman \& Begelman 1987; Goodman 2003). While many authors proposed possible solutions to this problem (e.g., Goodman 2003), it seems such a gravitationally  unstable disc is a good explanation for the existence of the first starts in galactic centers or the formation of super massive stars in quasar accretion discs (Goodman \& Tan 2004; Tan \& Blackman 2005; Levin  2007).  However, as far as we know, the effects of outflow on the gravitational stability of accretion discs of ANGs or around supermassive black holes have been neglected for simplicity. Many authors have studied the gravitational stability of discs without outflows during recent years. For example, Goodman \& Tan (2003)  estimated the mass of first clumps in quasar discs. Assuming that the viscosity is proportional to gas pressure, they found that the self-gravitating radius is ranging from $1700$ to $2700$ Schwarzschild radius, with a few hundred solar mass for the fragments. In another study, Nayakshin (2006) studied star formation near to our Galactic center. The mass of the first stars of this model is a few solar mass which may increase because of the subsequent accretion. Also,  Khajenabi \& Shadmehri (2007) studied  accretion discs around quasi-stellar objects (QSOs) and the Galactic center with a corona and determined the self-gravitating radius and the mass of the first clumps. They showed that  existence of a corona implies a more gravitationally unstable disc.

The purpose of the present work is to study possible effects of outflows on the gravitational stability of accretion discs around supermassive black holes. We present a set of steady-state analytical solution for the structure of an accretion disc with outflow. Basic assumptions and the main equations are presented in the next section. Gravitational stability of the disc is analyzed in section 3 by determining  the self-gravitating radius and the mass of the first clumps. Summary of the astrophysical implications is given in section 4.

\section{General Formulation}

We are following an approach similar to Knigge (1999) in order to include outflows in the  main equations. We can start by defining the cumulative mass-loss rate from the disc as
\begin{equation}
\dot M_{\rm w}(R)=\int_{R_{\rm in}}^{R} \dot m_{\rm w}(R') 4\pi R' dR',
\end{equation}
where $R_{\rm in}$ denotes radius at the inner edge of the disc and $\dot m_{\rm w}(R)$ is the mass-lose rate per unit area from each disc face. Having  this definition, the total mass loss rate is given by

\begin{equation}
\dot M_{\rm w,total}(R)= \dot M_{\rm w}(R_{\rm disc}),
\end{equation}
where $R_{\rm disc}$ is the radius of the accretion disc.

The equation of mass conservation is written as
\begin{equation}\label{eq:conserve}
\frac{\partial \Sigma}{\partial t}=\frac{-1}{R}\frac{\partial}{\partial R}(R \Sigma V_{\rm R})-\frac{1}{2\pi R}\frac{\partial \dot M_{\rm w}}{\partial R},
\end{equation}
where $V_{\rm R}$ is the radial inflow velocity of material in the disc $(V_{\rm R}<0)$. Note that the last term on right hand side of equation of mass conservation represents mass loss by the outflow.

Also, we can write equation of conservation of angular momentum,
\begin{displaymath}
\frac{\partial }{\partial t}(\Sigma R^{2} \Omega)=\frac{-1}{R}\frac{\partial}{\partial R}(R^{3}\Sigma V_{\rm R} \Omega )+ \frac{1}{R}\frac{\partial}{\partial R}(R^{2} \tau_{\rm r\varphi})
\end{displaymath}
\begin{equation}
-\frac{(lR)^{2}\Omega}{2\pi R}\frac{\partial \dot M_{\rm w}}{\partial R} \label{eq:pidirec}.
\end{equation}
The first two terms on the right-hand side of this equation describe the inflow of the angular momentum through the boundaries of the annulus and the effects of viscous torques due to shear (here, $\tau_{\rm r\phi}$ is viscous stress). The third term allows for the angular momentum sink provided by the outflow. Parameter $l$ permits the most meaningful division of the available parameter space and, in principle, allows different types of accretion disc winds to be identified. Choice of $l=1$ should be most appropriate for radiation-driven outflows (e.g. Murray \& Chiang 1996; Proga et al. 1998) and $l>1$ is appropriate for centrifugally driven MHD winds (e.g. Blandford \& Payne 1982; Pelletier \& Pudritz 1992).

Vertical hydrostatic equilibrium of the disc implies
\begin{equation}
\frac{p}{\Sigma}=\frac{\Omega_{\rm K}^2 H}{2}\label{eq:zidirec},
\end{equation}
and to keep the model simple, the disc will now be assumed to be Keplerian and in a steady state. The first of these assumptions gives
 \begin{equation}
\Omega(R)=(\frac{GM}{R^{3}})^{1/2},
\end{equation}
where $M$ is the mass of the central object, whereas the second implies that $\partial/\partial t\equiv 0$ in (\ref{eq:conserve}) and (\ref{eq:pidirec}). The conservation equation (\ref{eq:conserve}) can then be integrated as
 \begin{equation}\label{eq:cons2}
-2\pi R \Sigma V_{\rm R}-\dot M_{\rm w}(R)=constant.
\end{equation}
The first term on the left-hand side of this expression is clearly the accretion rate at radius $R$, i.e.  $\dot M_{\rm acc}(R)$.   The constant of integration in  equation (\ref{eq:cons2}) can be found by considering the boundary condition at  $R=R_{\rm in}$, which shows that it is simply the rate of accretion on to the central object, i.e.
 \begin{equation}
\dot M_{\rm acc}(R)=\dot M_{\rm acc}(R_{\rm in})+\dot M_{\rm w}(R).
\end{equation}
The angular momentum integral is somewhat more difficult to compute, since a result of the factor $R^{2}\Omega$ in the new outflow sink term, the right-hand side of equation (\ref{eq:pidirec}) is no longer a perfect differential (Knigge 1999). That term , at least, must therefore be integrated explicitly, which requires $\dot M_{\rm w}(R)$, or equivalently, $\dot m_{\rm w}(R)$ to be specified.

A simple power-law profile has been used by many authors (e.g.,  Quataert \& Narayan 1999; Knigge 1999; Beckert 2000; Turolla \& Dullemond 2000; Misra \& Taam 2001; Fukue 2004).  So,
 \begin{equation}
\dot m_{\rm w}(R)=K R^{\xi},\label{eq:m}
\end{equation}
where the constant $K$ being fixed by normalizing to $\dot M_{\rm w,total}$ (which is used as a free parameter) and $\xi$ that is the radial mass-loss power-law index which is used as a free parameter also.

A more general prescription for the viscous stress $\tau_{\rm r\phi}$ is considering (Taam \& Lin 1984; Watarai \& Mineshige 2003; Merloni \&  Nayakshin 2006):
\begin{equation}
\tau_{\rm r\phi}=-\alpha_{0} p^{1-\mu/2} p_{\rm gas}^{\mu/2},\label{eq:visg}
\end{equation}
where $\alpha_{0}$ and $0\leq\mu\leq 2$ are constants and $p$ is the sum of the gas and radiation pressures. Also, $\beta$ is  the ratio of gas pressure to the total pressure.

Having equation (\ref{eq:visg}) as prescription for the viscous stress and using simple power-law form for
$\dot m_{\rm w}(R)$, equation (\ref{eq:m}), then equation (\ref{eq:pidirec}) can be integrated straightforwardly. After mathematical manipulations, we can write equation (\ref{eq:pidirec}) as
\begin{displaymath}\
8\pi\alpha_{0} H (p_{\rm gas})^{\mu/2} p^{(2-\mu)/2}= 3\Omega_{\rm K} \dot{M}_{\rm acc}(R_{\rm disc})
\end{displaymath}
\begin{equation}
\times J(R)\left [ 1-\Psi(R) \right ] \label{eq:pi},
\end{equation}
where,
\begin{displaymath}
\Psi(R)=\frac{\dot M_{\rm w,total}}{\dot M_{\rm acc}(R_{\rm disc})}\frac{\chi(R)}{J(R)},
\end{displaymath}
\begin{displaymath}
J(R)=[1-(\frac{R_{\rm in}}{R})^{1/2}],
\end{displaymath}
and
\[\chi(R) = \left\{
\begin{array}{l l}
\vspace{0.1cm}
  $$1-(R_{\rm in}/R)^{1/2}-\left [\displaystyle \frac{(R/R_{\rm in})^{\xi+2}-1}{(R_{\rm disc}/R_{\rm in})^{\xi+2}-1} \right ]
$$\\

\vspace{0.1cm}

 -(R_{\rm in}/R)^{1/2} \left [l^{2}(\xi+2)/(\xi+5/2) \right ] \\

 \vspace{0.4cm}

 \times \left [\displaystyle \frac{1-(R/R_{\rm in})^{\xi+5/2}}{(R_{\rm disc}/R_{\rm in})^{\xi+2}-1} \right ], \hspace{0.5cm} $$\xi \neq -2, -5/2$$\\

\vspace{0.1cm}

  $$1-(R_{\rm in}/R)^{1/2}-\left [\displaystyle \frac{\ln (R/R_{\rm in})}{\ln (R_{\rm disc}/R_{\rm in})} \right ]\\

  \vspace{0.1cm}

  - \left [\displaystyle \frac{1-(R/R_{\rm in})^{1/2}}{\ln(R_{\rm disc}/R_{\rm in})} \right ]$$\\

  \vspace{0.4cm}

  $$\times 2l^{2}(R_{\rm in}/R)^{1/2}$$, \hspace{1.4cm} \quad \mbox{$\xi = -2$}\\

 \vspace{0.1cm}

  $$1-(R_{\rm in}/R)^{1/2}-\left [\displaystyle \frac{(R/R_{\rm in})^{-1/2}-1}{(R_{\rm disc}/R_{\rm in})^{-1/2}-1} \right ]\\

  \vspace{0.1cm}

  - \left [\displaystyle \frac{\ln (R_{\rm in}/R)}{1-(R_{\rm in}/R_{\rm disc})^{1/2}} \right ]$$\\

  \times (l^{2}/2)(R_{\rm in}/R)^{1/2} , \hspace{1cm} \quad \mbox{$\xi = -5/2$}\\ \end{array} \right. \]

The energy balance of the disc is given by
\begin{equation}\label{eq:energy01}
D(R) = \sigma T_{\rm eff}^{4} + \epsilon_{\rm w}(R),
\end{equation}
where $D(R)$ is the viscous dissipation in the disc and $\epsilon_{\rm w}(R)$  represents energy-loss rate by the outflow. As in Knigge (1999), we parameterize the outflow loss term $\epsilon_{\rm w}$: the rate at which energy must be supplied to outflow in order for it to overcome its binding energy, i.e $\epsilon_{\rm w} = (1/2)\eta_{\rm b}\dot{m}_{\rm w} v_{\rm K}^{2}$, where $\eta_{\rm b} \leq 1$ is a efficiency constant. Also, kinetic energy has to be supplied to the outflow at a rate of $\epsilon_{\rm k}=(1/2)\eta_{\rm k}\dot{m}_{\rm w} v_{\infty}^{2}$, where $\eta_{\rm k} \leq 1$ is a efficiency constant and $v_{\infty}(R) = f v_{\rm K}$ ($f$ is a constant). Thus, equation (\ref{eq:energy01}) becomes
\begin{equation}\label{eq:energy02}
D(R) = \sigma T_{\rm eff}^{4} + \frac{1}{2}(\eta_{\rm b}+\eta_{\rm k}f^{1})\dot{m}_{\rm w}v_{\rm k}^{2}(R),
\end{equation}
and finally after mathematical manipulation, the energy equation (\ref{eq:energy02}) is written as
\begin{equation}
\sigma T_{\rm eff}^{4}=\frac{3}{8\pi}\Omega_{\rm K}^{2}\dot M_{\rm acc}(R_{\rm disc}) J(R)(1-\Psi(R)-\Theta(R))\label{eq:energy},
\end{equation}
where
\begin{equation}
\Theta(R)=\frac{4\pi}{3}(\eta_{\rm b}+\eta_{\rm k}f^{2})\frac{R^{2}\dot m_{\rm w}(R)} {\dot{M}_{\rm acc}(R_{\rm disc})}
\frac{1}{J(R)},
\end{equation}
or
\[\Theta(R) = \left\{
\begin{array}{l l}
\vspace{0.1cm}
  $$\displaystyle \frac{1}{3}(\eta_{\rm b}+\eta_{\rm k}f^{2})\displaystyle\frac{\dot{M}_{\rm w, total}}{\dot{M}_{\rm acc}(R_{\rm disc})}\displaystyle \frac{(R/R_{\rm in})^{\xi +2}}{J(R)}$$\\

\vspace{0.1cm}

 \times \displaystyle [ \frac{\xi +2}{(R_{\rm disc}/R_{\rm in})^{\xi+2}-1}], \hspace{0.5cm} $$\xi \neq -2, -5/2$$ \\

\vspace{0.1cm}

  $$\displaystyle \frac{1}{3}(\eta_{\rm b}+\eta_{\rm k}f^{2})\displaystyle\frac{\dot{M}_{\rm w, total}}{\dot{M}_{\rm acc}(R_{\rm disc})}\frac{1}{\ln (R_{\rm disc}/R_{\rm in})}\\

  \vspace{0.1cm}

  $$\times \displaystyle \frac{1}{J(R)}$$, \hspace{2.2cm} \quad \mbox{$\xi = -2$}\\

 \vspace{0.1cm}

  $$\displaystyle \frac{1}{6}(\eta_{\rm b}+\eta_{\rm k}f^{2})\displaystyle\frac{\dot{M}_{\rm w, total}}{\dot{M}_{\rm acc}(R_{\rm disc})}\frac{1}{1- (R_{\rm in}/R_{\rm disc})^{1/2}}\\

  \vspace{0.1cm}

  $$\displaystyle \times \frac{(R_{\rm in}/R)^{1/2}}{J(R)}$$ . \hspace{1.4cm} \quad \mbox{$\xi = -5/2$}\\ \end{array} \right. \]

Finally, with the vertical transport of heat dominated by radiative diffusion
we have a relation between the midplane and surface temperatures which is given by
\begin{equation}
T = (\frac{3}{8}\kappa\Sigma)^{1/4} T_{\rm eff},\label{eq: T}
\end{equation}
where $\kappa$ is the opacity coefficient.

Equations (\ref{eq:zidirec}), ( \ref{eq:pi}), (\ref{eq:energy}) and
(\ref{eq: T}) enable us to find $p$
and $T$ and $\rho$ as functions of $R$ with $\beta$  the critical input
parameters. Thus,

\begin{displaymath}
T=(\frac{4\sigma \Omega_{\rm K}}{3\kappa \alpha_{0}})^{-1/2}(\frac{16\pi^{2}\alpha_{0}^{2}ck_{\rm B}}
{3\sigma \mu_{\rm m} m_{\rm H} \dot{M}_{\rm acc}(R_{\rm disc})^{2}\Omega_{\rm K}^{4}J^{2}})^{-1/3}
\end{displaymath}
\begin{equation}\label{eq:main1}
\times {(1-\beta)^{-1/3}}\beta^{(4-\mu)/12}(1-\Psi)^{1/6}(1-\Psi-\Theta)^{1/2},
\end{equation}
\begin{displaymath}
p=(\frac{4\sigma \Omega_{\rm K}}{3\kappa \alpha_{0}})^{-1/2}(\frac{16\pi^{2}\alpha_{0}^{2}ck_{\rm B}}
{3\sigma \mu_{\rm m} m_{\rm H} \dot{M}_{\rm acc}(R_{\rm disc})^{2}\Omega_{\rm K}^{4}J^{2}})^{-2/3}
\end{displaymath}
\begin{equation}\label{eq:main2}
\times(1-\beta)^{-2/3}\beta^{(8-5\mu)/12}(1-\Psi)^{5/6}(1-\Psi-\Theta)^{1/2},
\end{equation}
\begin{equation}
\rho=(\frac{8\pi \alpha_{0}}{3\Omega_{\rm K}^{2}\dot{M}_{\rm acc}(R_{\rm disc})J(R)})^{2}p^{3}\beta^{\mu}(1-\Psi)^{-2}.
\end{equation}
There is an algebraic equation  for  $\beta$ as follows
\begin{displaymath}
 (\frac{4\sigma\Omega_{\rm K}}{3\kappa \alpha_{0}})^{-3/2}(\frac{8\pi \alpha_{0}}{3\Omega_{\rm K}^{2}\dot{M} J})^{2} (\frac{16\pi^{2}\alpha_{0}^{2}c k_{\rm B}}{3\sigma\mu_{\rm m} m_{\rm H} \dot{M}^{2}\Omega_{\rm K}^{4} J^{2}})^{-5/3}
\end{displaymath}
\begin{equation}
 \times\frac{k_{\rm B}}{\mu_{\rm m} m_{\rm H}} (1-\beta)^{-5/3}\beta^{(8+\mu)/12}(1-\Psi)^{-1/6}(1-\Psi-\Theta)^{3/2}=1,\label{eq:beta}
\end{equation}
where $\mu_{\rm m}$ is the mean particle mass in units of the hydrogen atom mass, $m_{\rm H}$. The other constants have their usual meanings.
In order to study the behavior of our solutions, it is more convenient
to introduce dimensionless variables. For the central mass $M$,  we
introduce $M_8=M/(10^{8}M_{\odot})$ and for the radial distance $R$,
we have $r_3=R/(10^{3}R_{\rm S})$, where $R_{\rm S}=2GM/c^{2}$ is
the Schwarzschild radius. The mass accretion rate can be written as
\begin{equation}
\dot{M}_{\rm acc}(R_{\rm disc})=\frac{l_{\rm E}}{\epsilon}\frac{4\pi GM}{\kappa_{\rm e.s.}c}=\frac{l_{\rm E}}{\epsilon}\frac{L_{\rm E}}{c^{2}},
\end{equation}
where $l_{\rm E}=L/L_{\rm E}$ is the dimensionless disc luminosity relative to the Eddington limit,
$\epsilon=L/(\dot{M}c^{2})$ is the radiative efficiency and $\kappa_{\rm e.s.}\approx 0.04$
$\rm m^{2} kg^{-1}$ is the electron opacity.

\begin{figure*}
\vspace*{-150pt}
\includegraphics[scale=0.9]{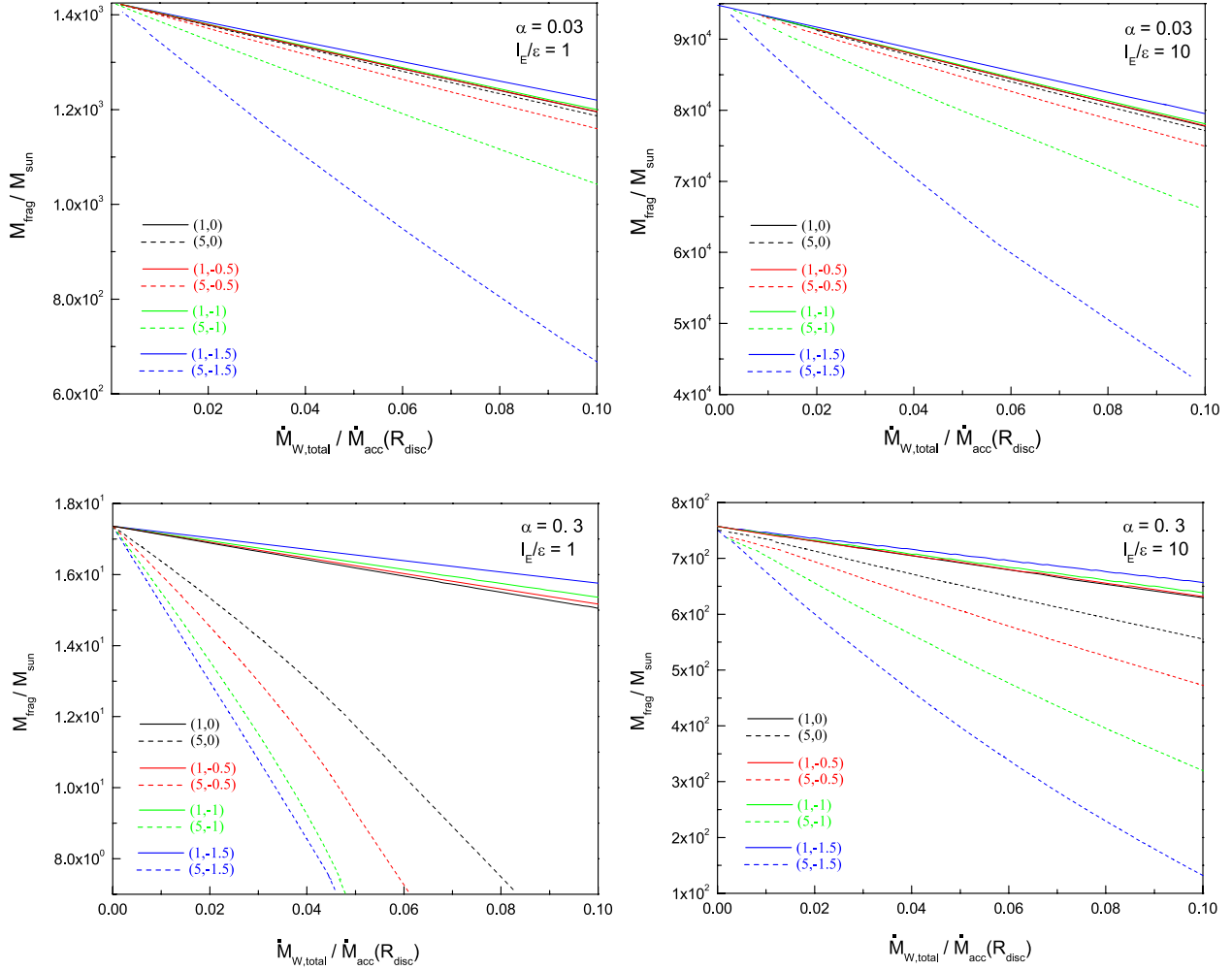}
\caption{Mass of the first clumps, $M_{\rm frag}$, (in  solar mass) at the self-gravitating radius versus the ratio of the total mass-loss by the outflow to the accretion rate, $\dot{M}_{\rm w, total}/\dot{M}_{\rm acc}(R_{\rm disc})$, for $\mu=2.0$, $\mu_{\rm m}=0.6$, $\hat{k}=1$, $M_{8}=1$, $f=\sqrt{2}$,   $\eta_{\rm b}=1$ and $\eta_{\rm k}=1$. Solid and dashed curves are corresponding to the solutions with $l=1$ and $l=5$, respectively. Each curve is marked by a pair $(l, \xi)$. The other input parameter are $\alpha=0.03$ and $l_{\rm E}/\epsilon = 1$ (top, left), $\alpha=0.03$ and $l_{\rm E}/\epsilon = 10$ (top, right), $\alpha=0.3$ and $l_{\rm E}/\epsilon = 1$ (bottom, left), $\alpha=0.3$ and $l_{\rm E}/\epsilon = 10$ (bottom, right). The general treatment of the solutions shows that as more mass is extracted from the disc by outflow, i.e. a stronger outflow, the first clumps are forming with smaller mass.}\label{fig:f1}
\end{figure*}
\begin{figure*}
\vspace*{-150pt}
\includegraphics[scale=0.9]{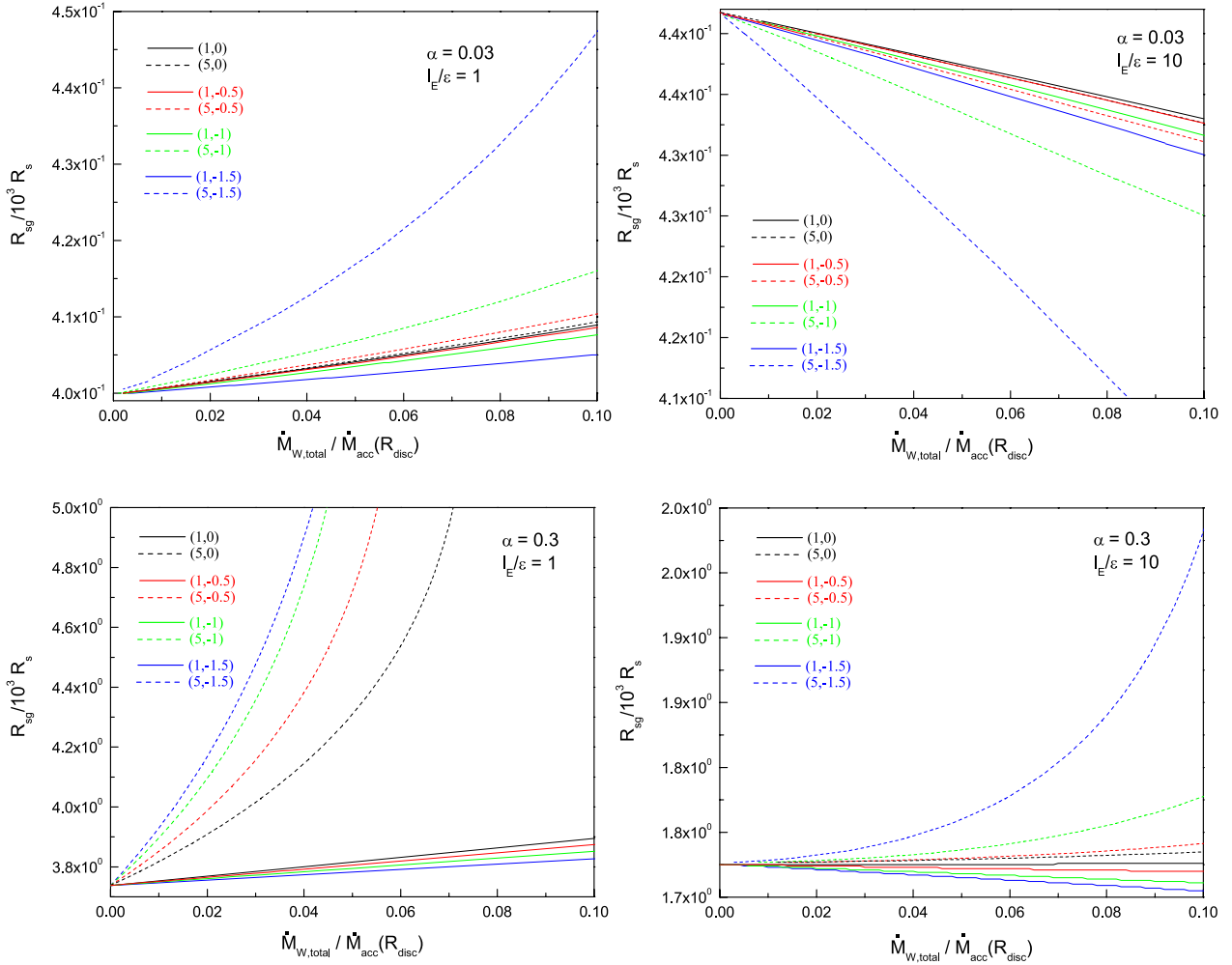}
\caption{The same as Figure \ref{fig:f1}, but it shows self-gravitating radius, $R_{\rm sg}$, (in Schwarzschild radius $R_{\rm S}$) versus the ratio of the total mass-loss by the outflow to the accretion rate, $\dot{M}_{\rm w, total}/\dot{M}_{\rm acc}(R_{\rm disc})$. Except for the case of small viscosity coefficient and high accretion rate, i.e. $\alpha=0.03$ and $l_{\rm E}/\epsilon = 10$, generally, self-gravitating radius increases as the outflow becomes stronger.}\label{fig:f2}
\end{figure*}

Our solution for $T$, $P$ and $\rho$ are written as
\begin{displaymath}
 \rho=2.76\times10^{-6} \alpha_{0}^{-1/2}\hat{\kappa}^{3/2} M_{8}^{-1/2}(\frac{l_{\rm E}}{\epsilon})^{2} J^{2} r_{3}^{-15/4}
\end{displaymath}
\begin{equation}
\times \beta^{(8-\mu)/4}(1-\beta)^{-2}(1-\Psi)^{1/2}(1-\Psi-\Theta)^{3/2},\label{eq:rho}
\end{equation}

\begin{displaymath}
p=27.15\alpha_{0}^{-5/6}\hat{\kappa}^{1/2} M_{8}^{-5/6}(\frac{l_{\rm E}}{\epsilon})^{4/3} J^{4/3} r_{3}^{-13/4}
\end{displaymath}
\begin{equation}
 \times \beta^{(8-5\mu)/12}(1-\beta)^{-2/3}(1-\Psi)^{5/6}(1-\Psi-\Theta)^{1/2},\label{eq:p}
\end{equation}

\begin{displaymath}
T=0.714\times10^{3}\alpha_{0}^{-1/3}\hat{\kappa}^{-1} M_{8}^{-1/3}(\frac{l_{\rm E}}{\epsilon})^{-2/3} J^{-2/3} r_{3}^{1/2}
\end{displaymath}
\begin{equation}
 \times \beta^{-(2+\mu)/6}(1-\beta)^{4/3}(1-\Psi)^{1/3}(1-\Psi-\Theta)^{-1},\label{eq:p}
\end{equation}
\begin{displaymath}
 \frac{H}{R}=4.67\times10^{-3}\alpha_{0}^{-1/6}\hat{\kappa}^{-1/2} M_{8}^{-1/6}(\frac{l_{\rm E}}{\epsilon})^{-1/3} J^{-1/3}
\end{displaymath}
\begin{equation}
 \times \beta^{-(8+\mu)/12}(1-\beta)^{2/3}(1-\Psi)^{1/6}(1-\Psi-\Theta)^{-1/2},\label{eq:HR}
\end{equation}
and the ratio $\beta$ is obtained from nondimensional form of equation (\ref{eq:beta}), i.e.
\begin{displaymath}
 0.16 \alpha_{0}^{1/6}\hat{\kappa}^{3/2} M_{8}^{1/6}(\frac{l_{\rm E}}{\epsilon})^{4/3} J^{4/3} r_{3}^{-7/4}
\end{displaymath}
\begin{equation}
\times \beta^{(8+\mu)/12}(1-\beta)^{-5/3}(1-\Psi)^{-1/6}(1-\Psi-\Theta)^{3/2}=1,\label{eq:beta1}
\end{equation}
where $\hat{\kappa}=\kappa/\kappa_{\rm e.s.}$
 We can also calculate the surface density as
\begin{displaymath}
\Sigma = 1.27\times 10^{5} \alpha_{0}^{-2/3}\hat{\kappa} M_{8}^{1/3}(\frac{l_{\rm E}}{\epsilon})^{5/3} J^{5/3} r_{3}^{-11/4}
\end{displaymath}
\begin{equation}
\times \beta^{(4-\mu)/3}(1-\beta)^{-4/3}(1-\Psi)^{2/3}(1-\Psi-\Theta).
\end{equation}

Toomre (1964) showed that a rotating disc is subject to
gravitational instabilities when the $Q$-parameter becomes smaller
than a critical value, which is close to unity,
\begin{equation}
Q=\frac{c_{\rm s} \Omega}{\pi G \Sigma},
\end{equation}
where $c_{\rm s}$ is the sound speed inside the accretion disc and $\Omega = \Omega_{\rm K}$ is the angular velocity. So, the Toomre parameter
of our model becomes
\begin{displaymath}
Q = 44 \alpha_{0}^{1/2}\hat{\kappa}^{-3/2} M_{8}^{-3/2}(\frac{l_{\rm E}}{\epsilon})^{-2} J^{-2} r_{3}^{3/2}
\end{displaymath}
\begin{equation}
\times \beta^{-(8-\mu)/4}(1-\beta)^{2}(1-\Psi)^{-1/2}(1-\Psi-\Theta)^{-3/2}.\label{eq:Toomrem}
\end{equation}

This equation with algebraic equation (\ref{eq:beta1}) gives the Toomre parameter as a function of the radial distance. Note that in the case of no-outflow our solutions reduce to standard disc solutions. Generally, the Toomre parameter is much higher than unity in the inner parts of the disc which implies these regions are gravitationally stable and do not fragment (e.g., Khajenabi \& Shadmehri 2007). But the Toomre parameter decreases with increasing  radial distance so that $Q$ reaches the critical value of unity at a self-gravitating radius which we denote  by $R_{\rm  sg}$. Thus, all regions with $R>R_{\rm sg}$ are gravitational unstable and may fragment to clumps and cores.

Different authors estimate the mass of fragments differently. Since
the disc is marginally unstable, the initial sizes and masses of
gravitationally bound fragments can be determined by Toomre's
dynamical instability (Toomre 1964). The most unstable wavelength for the $Q \sim
1$ disc is of order of the disc vertical scale height $H$ (Toomre
1964). Thus, the most unstable mode has radial wave number $k_{\rm
mu}=(QH)^{-1}$ and so the mass of a fragment at $R=R_{\rm sg}$
becomes
\begin{equation}
M_{\rm frag} \approx \Sigma (\frac{2\pi}{k_{\rm mu}})^{2}=4\pi^{2}\Sigma H^{2}.\label{eq:frag}
\end{equation}
In the next section, we calculate the self-gravitating radius and the mass of the first clumps according to our analytical solutions.
\section{Analysis}

\begin{figure*}
\includegraphics{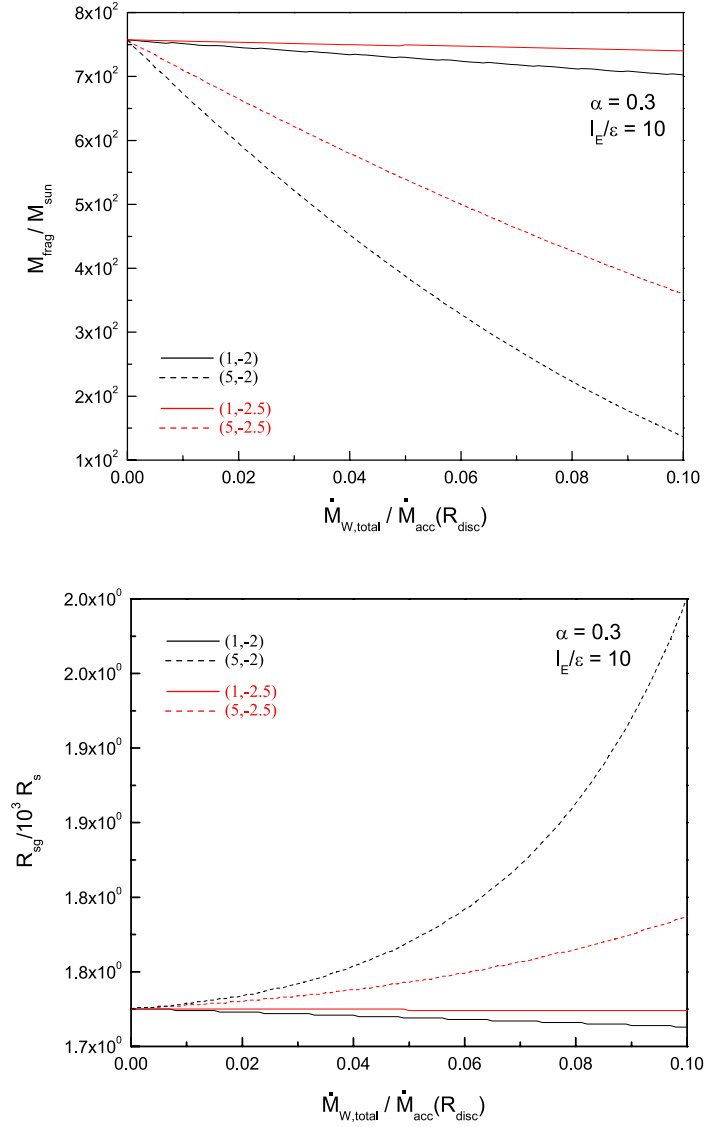}
\caption{The self-gravitating radius, $R_{\rm sg}$, (in Schwarzschild radius $R_{\rm S}$) and the mass of the first clumps, $M_{\rm frag}$, (in  solar mass) versus the ratio of the total mass-loss by the outflow to the accretion rate, $\dot{M}_{\rm w, total}/\dot{M}_{\rm acc}(R_{\rm disc})$, for $\mu=2.0$, $\mu_{\rm m}=0.6$, $\hat{k}=1$, $M_{8}=1$,  $f=\sqrt{2}$, $\eta_{\rm b}=1$ and $\eta_{\rm k}=1$. Here, the index of mass-loss rate by outflows is adopted as $\xi = -2$ and $-2.5$. Solid and dashed curves are corresponding to the solutions with $l=1$ and $l=5$, respectively. Each curve is marked by a pair $(l, \xi)$. }\label{fig:f3}
\end{figure*}

We fix the central mass, the opacity  and
the mean molecular weight, respectively as $M_{8}=1$,
$ \hat{\kappa}=1$  and  $\mu_{m}=0.6$. Also we have $f=\sqrt{2}$, $\eta_{\rm b}=1$, $\eta_{\rm k}=1$ and  $ \mu=2 $. Since for the other input parameters we find similar qualitative results, in order to avoid reputation, our analysis is restricted to the mentioned input parameters. We find that variations of the mass of fragments with the  ratio $\dot{M}_{\rm w, total}/\dot{M}_{\rm acc}(R_{\rm disc})$ is more sensitive to the higher values of the exponent of viscosity, $ \mu $,  and so we adopt  $ \mu=2 $ in our plots. Also, the ratio of gas pressure and total pressure, $\beta$, is not very sensitive to the ratio of the total mass-loss rate and the accretion rate  for different input parameters. The other input parameters related to the outflow are changed to illustrate their possible effects on the physical properties of the system. The mass-loss power-law index $\xi$ is adopted values of $0$, $-0.5$, $-1$ and $-1.5$ in Figures \ref{fig:f1} and \ref{fig:f2} and  also, we have $\xi=-2$  and $-2.5$ in Figure \ref{fig:f3}.

We also consider $l=1$ and $5$. As noted, radiation driven disc winds are expected to belong to the $l=1$ family of models and  centrifugally driven  disc winds are expected to belong to the $l=5$ family of models.  In our analysis, we will use the nondimensional factor  $l_{\rm E}/\epsilon$ as a free parameter so that by changing this parameter we can consider  appropriate values of the accretion rate. Evidently, higher the ratio $l_{\rm E}/\epsilon$, larger the accretion rate. However, some authors introduce different values for the accretion rate. For example, for a central mass with mass $M=10^{8} M_{\odot}$, Goodman \& Tan (2004) proposed $l_{\rm E}/\epsilon = 10$. Thus, in our analysis, the chosen values of $l_{\rm E}/\epsilon =1$ and $10$ are acceptable, also we are using  values of $0.03$ and $0.3$ for $\alpha_{0}$.

 Using equation (\ref{eq:frag}), we can calculate the mass of the first fragments for the set of the above input parameters. Figure \ref{fig:f1} shows  the mass of the fragments  at the self-gravitating radius versus the  ratio of the total mass loss rate and the accretion rate  for different input parameters. Each curve is represented by a pair of  $l$ and the mass-loss power-low index $\xi$ as $(l,\xi)$.
 For a given set of the input parameters, a higher accretion rate implies  fragments with higher mass. For example, while for a case with $\alpha_{0}=0.03$ and $l_{\rm E}/\epsilon = 1$, the mass of the first clumps will be between $6\times 10^2 \rm M_{\odot}$ and $1.4\times 10^3 \rm M_{\odot}$ depending on mass-loss rate by outflow, we see that for $\alpha_{0}=0.03$ and $l_{\rm E}/\epsilon = 10$ the mass of fragments will increase up to  a value between $4\times10^{4} \rm M_{\odot}$ and $9\times10^{4} \rm M_{\odot}$. Also, for a fixed accretion rate, the mass of the first clumps decreases with increasing $\alpha_{0}$.  In all plots of Figure \ref{fig:f1} for $l=5$, we see that mass of fragments is  more sensitive to the variations of  the mass-loss power-law index $\xi$ in comparison to the solutions with $l=1$. In other words, when outflows are centrifugally driven, the mass of the fragments highly depends on the mass-loss index $\xi$. But for radiation driven outflows this dependence is weak. Also, in the case of centrifugally driven outflows, for a fixed ratio $\dot{M}_{\rm w, total}/\dot{M}_{\rm acc}(R_{\rm disc})$  the mass of fragments increases with increasing the mass-loss index $\xi$. But in the case of radiation driven outflows we see an opposite variation, i.e. the mass of fragments decreases with increasing the mass-loss index $\xi$ if the other input parameters are kept fixed.
 Also our plots show that for all input parameter when the ratio  $\dot{M}_{\rm w, total}/\dot{M}_{\rm acc}(R_{\rm disc})$ increases, then  mass of the first clumps decreases and interestingly the reduction is linearly proportional to the mass-loss rate, i.e. $M_{\rm frag} \cong - A(\dot{M}_{\rm w, total}/\dot{M}_{\rm acc}(R_{\rm disc}))+B$ where constants $A$ and $B$ depend on the input parameters.

 Figure \ref{fig:f2} shows  the self-gravitating radius $R_{\rm sg}$  (in Schwarzschild radius $R_{\rm S}$) versus the  ratio of the total mass loss rate and the accretion rate  for different input parameters. For a given set of the input parameters, the self-gravitating radius increases with increasing  the viscosity coefficient $\alpha_{0}$. The self-gravitating radius  is not very  sensitive to the variations of  the mass-loss  index $\xi$ when there are radiation driven outflows. When there are centrifugally driven outflows, the self-gravitating radius increases with decreasing the mass-loss  index, except for a case with small viscosity coefficient and high accretion rate (i.e., $\alpha_{0}=0.03$ and $l_{\rm E}/\epsilon = 10$). Also,  the self-gravitating radius decreases with increasing mass-loss rate by the outflow unless the viscosity coefficient is small and the accretion rate is high. In the case of $\xi=-2$ and $-2.5$, we can also determine self-gravitating radius and the mass of the fragments. Figure \ref{fig:f3} shows the mass of the fragments (top) and the self-gravitating radius $R_{\rm sg}$ (bottom) versus the  ratio of the total mass loss rate and the accretion rate. Variations of the mass of the fragments and self-gravitating radius with the input parameters are similar to Figures \ref{fig:f1} and \ref{fig:f2}.

\section{Conclusion}
We presented a set of analytical solutions for the steady state structure of discs with outflows around supermassive black holes. Gravitational stability of the disc has been studied using our analytical solutions. We determined the self-gravitating radius and the  mass of the first clumps at this radius. We showed that as more mass, angular momentum and energy are extracted from the disc by the outflows, the mass of the first fragments decreases, though the self-gravitating radius increases except for a case with small viscosity coefficient and high accretion rate. However,  we think low values of $\alpha_{0}$ are not acceptable in self-gravitating discs. We can conclude that the existence of outflows imply  a more gravitationally stable accretion disc. Interestingly, there is a linear correlation between the mass of the fragments and mass-loss rate by the outflows, according to our solutions. The mentioned effects of the outflows on the gravitational stability of the discs should be considered in theoretical studies of star formation near to the supermassive black holes.

\acknowledgments

I gratefully acknowledge Peter Duffy for his support and encouragement. I am grateful for Ad Astra PhD Scholarship of University College of Dublin.

\end{document}